\documentclass[aps,prb,floatfix,twocolumn,showpacs,10pt]{revtex4-1}
\usepackage{graphicx}
\usepackage{amsmath}
\usepackage{amssymb}
\usepackage{amsfonts}

\usepackage[usenames,dvipsnames]{color}

\graphicspath {{Figures/}}

\begin{document}

\title{Optimal geometry of lateral GaAs and Si/SiGe quantum dots for electrical control of spin qubits}

\begin{abstract}
We investigate the effects of the orientation of the magnetic field and the orientation of a quantum dot, with respect to crystallographic coordinates, on the quality of an electrically controlled qubit realized in a gated semiconductor quantum dot. We find that, due to the anisotropy of the spin-orbit interactions, by varying the two orientations it is possible to tune the qubit in the sense of optimizing the ratio of its couplings to phonons and to a control electric field. We find conditions under which such optimal setup can be reached by solely reorienting the magnetic field, and when a specific positioning of the dot is required. We also find that the knowledge of the relative sign of the spin-orbit interaction strengths allows to choose  a robust optimal dot geometry, with the dot main axis along [110], or [1$\overline{1}$0], where the qubit can be always optimized by reorienting the magnetic field.
\end{abstract}
\author{Ognjen Malkoc,$^{1,2}$ Peter Stano,$^{2,3}$ Daniel Loss$^{2,4}$}
\affiliation{$^1$Department of Physics, Lund University, Box 118, SE-221 00 Lund, Sweden\\
$^2$RIKEN Center for Emergent Matter Science, 2-1 Hirosawa, Wako, Saitama 351-0198, Japan\\
$^3$Institute of Physics, Slovak Academy of Sciences, 845 11 Bratislava, Slovakia\\
$^4$Department of Physics, University of Basel, Klingelbergstrasse 82, CH-4056 Basel, Switzerland}
\pacs{76.30.-v , 81.07.Ta , 81.05.Ea , 81.05.Hd, 71.70.Ej, 72.10.Di}

\maketitle

\section{Introduction}
Nanoscale devices provide a promising venue for solid state based quantum and classical information processing.  
Among prime candidates as information carriers are spins of localized electrons in quantum dots,\cite{loss1998:PRA} which combine high tunability with a weak coupling to the surrounding lattice.\cite{hanson2007:RMP,kloeffel2013:ARCMP} This architecture also presents a viable method for detecting and generating entangled pairs of electrons, an essential resource for quantum computation. \cite{malkoc2014:EPL,brange2015:PRL}
GaAs has been the prime material of choice for these structures,\cite{delbecq2014:APL,scarlino2014:PRL,yoneda2014:PRL,shulman2012:S,busl2013:NN,higginbotham2014:PRL} though Si has seen a lot of progress recently,\cite{maune2012:N,kawakami2014:NN, takeda2013:APL,shi2014:NC,hao2014:NC,veldhorst2015:N} motivated by compatibility with semiconductor industry and lower nuclear spin originated magnetic noise.\cite{chekhovich2013:NM}

Even though manipulating spins by an oscillating magnetic field (electron spin resonance) is conceptually most straightforward \cite{koppens2006:N}, technologically it is much more desirable to use electrical fields  [electric dipole spin resonance (EDSR) ] \cite{golovach2006:PRB,nowack2007:S,brunner2011:PRL}. One of the options \cite{golovach2006:PRB,rashba2008:PRB} is to exploit the spin-orbit coupling of the semiconductor \cite{zutic2004:RMP}.
With an applied magnetic field, the spin-orbit interaction mediates a coupling between the electron spin and the electric field. However, the same coupling makes the spin vulnerable to the electric noise of the environment, in addition to the magnetic noise \cite{kuhlmann2013:NP,koppens2007:PRL,chesi2015:PRL}. The latter is predominantly a low-frequency noise from nuclear spins, and can be efficiently suppressed by spin echo \cite{petta2005:S,bluhm2011:NP}, feedback control \cite{klauser2006:PRB,bluhm2010:PRL}, and a Hamiltonian estimation by fast measurements \cite{shulman2014:NC,delbecq2016:PRL}.
No such remedies are known for the high-frequency electrical noise, which then fundamentally limits the spin lifetime as predicted by theory \cite{khaetskii2001:PRB,woods2002:PRB,tahan2002:PRB,golovach2004:PRL} and confirmed in experiments \cite{fujisawa2002:N,hanson2003:PRL,meunier2007:PRL,amasha2008:PRL,xiao2010:PRL} .

\begin{figure}[ht]
\includegraphics[width=0.30\textwidth]{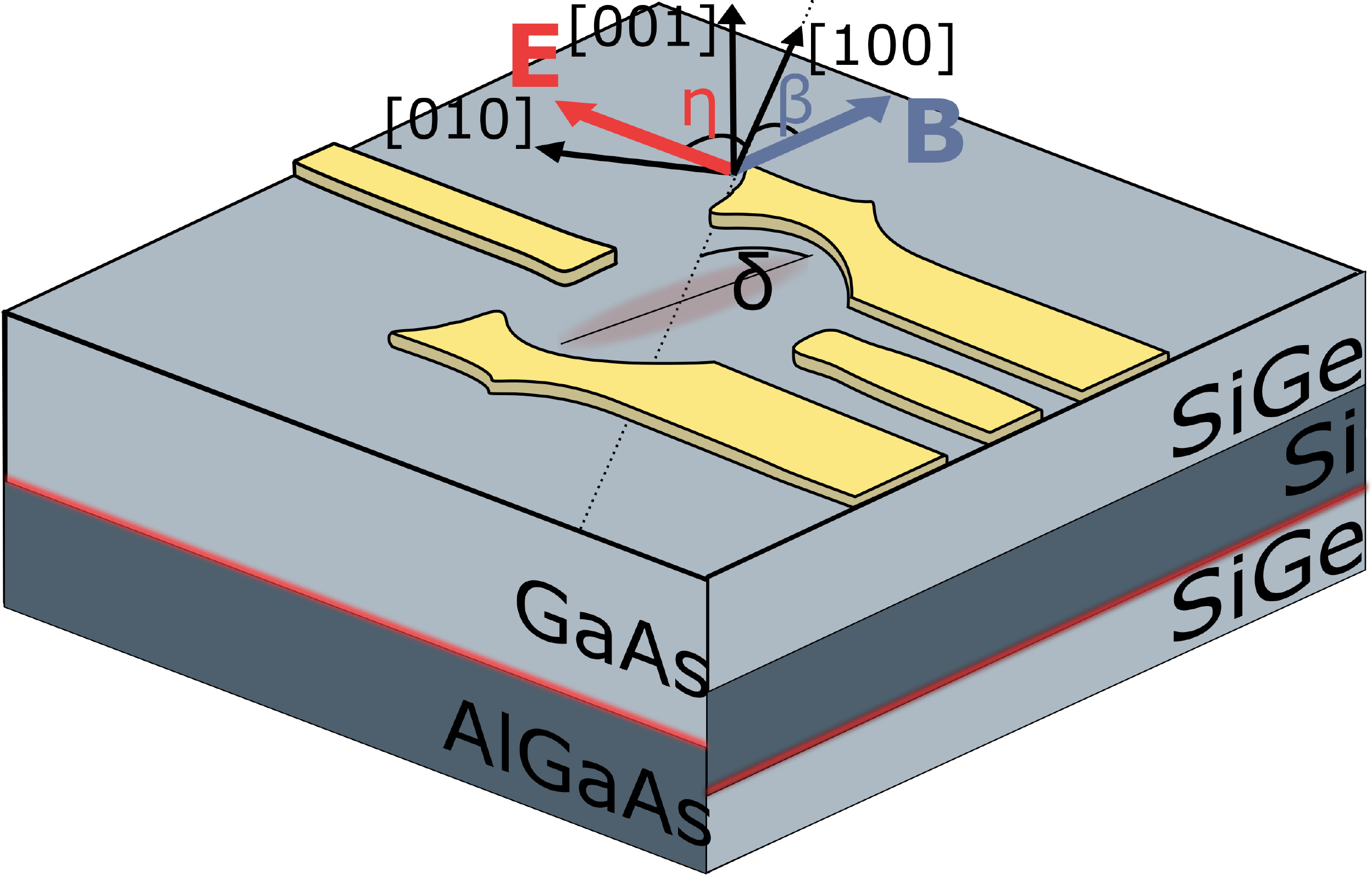}
\caption{A schematic illustration of the considered lateral quantum dots in the two cases of GaAs and Si/SiGe. An electron from the 2DEG at the interface, depicted as a thin red line, is trapped in an anharmonic trap potential with the trapping frequency $\omega_{x'} (\omega_{y'})$ along the major (minor) axis of the dot. The electron spin in the dot is controlled via Rabi oscillations, induced by the oscillating resonant electric field $\mathbf E$. The applied magnetic field $\mathbf B$ modulates the Rabi frequency and the spin relaxation rate.} 
\label{fig:schematic}
\end{figure}

Though there were some theoretical suggestions how to suppress the coupling of the spin to phonons in the presence of spin-orbit interactions, they require either parameter fine tuning \cite{climente2007:PRB,liao2006:PRB}, or a structure design \cite{chaney2007:PRB}, which are difficult to achieve experimentally. Here, we follow a much more robust way based on the following simple idea. The two dominant spin-orbit interactions in a two-dimensional electron gas (2DEG), the Rashba and Dresselhaus terms, have different dependence on the electron momentum in the 2DEG plane.
It has been realized early on that, similarly as in 2DEG \cite{averkiev1999:PRB}, it will lead to an anisotropic spin relaxation also in quantum dots \cite{rashba2003:PRL,golovach2004:PRL,bulaev2005:PRB,falko2005:PRL,stano2006:PRL,stano2006:PRB,olendski2007:PRB,golovach2008:PRB}, as  recently confirmed experimentally \cite{scarlino2014:PRL,nichol2015:NC}. Even though both the coupling to the phonons and to the electric control field vary due to the same anisotropy of spin-orbit interactions, the EDSR rate and the phonon induced relaxation rate are not proportional to each other. It then immediately follows that one can in principle optimize the ratio of these two rates upon changing the parameters under control: the direction of the magnetic field and the orientation of the dot within the crystallographic axes of the sample. 

What are the optimal dot geometries and how much can one gain by this optimization depending on the dot ellipticity and the ratio of spin-orbit strengths is the question which we investigate in this work. 
While Sec.~IV gives a detailed analysis, we pinpoint here some general features. First, the closer in strength the two spin-orbit interactions are and the more elliptic the dot confinement is, the more tunable is the coupling of the spin to the electric fields. This is because, respectively, the higher  the directional anisotropy of the spin-orbit interactions is,  the more precisely can the electron momentum be locked to a desired direction in the dot. On the other hand, if one of the spin-orbit interactions is dominant, only the relative orientation of the dot confinement and the magnetic field matters, and therefore the coupling is fully tunable by controlling only one of these directions. This is due to the rotational (quasi-rotational) symmetry of the Rashba (Dresselhaus) spin-orbit interaction. Finally, if the relative sign of the spin-orbit interactions is known, which seems easier to extract experimentally than the strengths themselves \cite{scarlino2014:PRL,raith2014:PSSB}, we show that a dot with the main axis along, depending on the sign, [110] or [1$\overline{1}$0], is a robust optimal design for experiments where a vector magnet is available.

\section{Spin qubit states}

Before we quantify and analyze its quality, we need to introduce the geometry and basic parameters of the quantum dot spin qubit, depicted schematically in Fig. ~\ref{fig:schematic}. To this end, we consider the Hamiltonian  
\begin{equation}
H = \frac{\mathbf{p}^2}{2 m^*} + V(x,y) + \frac{g^*\mu_B }{2} \boldsymbol\sigma \cdot \mathbf{B} + H_{so},
\label{eq:H}
\end{equation}
comprising, respectively, the kinetic energy, the confinement potential, the Zeeman term, and the spin-orbit interactions. In the quasi-two-dimensional description that we use, the electron position ${\bf r}$ and momentum $\mathbf{p} = i \hbar \nabla_{\bf r}$ are two-dimensional vectors. The effective electron  mass $m^*$ enters the kinetic energy term. The quantum dot confinement potential is taken as a bi-harmonic one, 
\begin{equation}
V(x,y)  = \frac{\hbar^2}{2 m^* } \left(\frac{{x'}^2}{l_{x'}^4} + \frac{{y'}^2}{l_{y'}^4} \right).
\label{eq:potential}
\end{equation}
The dot orientation is defined by the rotation angle $\delta$ relating the crystallographic coordinates $\mathbf r = (x,y)$ to the dot coordinates ${\bf r}^\prime=(x^\prime,y^\prime)$ through $x'= x\cos \delta  + y\sin \delta$, and $ y' = -x\sin \delta  +  y\cos \delta$. The dot shape is parametrized by the confinement length $l$ and ellipticity 
$\epsilon \in [ 0, 1 )$, 
where $l_{x'} = l$ and $l_{y'} = l (1-\epsilon)^{1/4}$. The two limiting cases correspond to, $\epsilon = 0$ for a circular dot, and $\epsilon \to 1$ for a purely one dimensional dot.  The magnetic field is considered to be in-plane to minimize the orbital effects, ${\bf B}=B(\cos \beta, \sin \beta,0) = B \bf{b}$. It then couples only to the electron spin, with the corresponding operator being the vector of Pauli matrices $\boldsymbol\sigma$. 

With these ingredients, the eigenstates of the Hamiltonian $H$ are separable into  orbital and spin parts, appearing in pairs with opposite spin. We denote $\sigma=\uparrow$ for the spinor parallel to the magnetic field and $\sigma=\downarrow$ for the anti-parallel one. The pair is split by the Zeeman energy $g^* \mu_B  B$, according to the electron $g$ factor $g^*$ and the Bohr magneton $\mu_B$.  Further, we consider temperatures much smaller than the orbital energy scale $k_BT \ll \hbar^2/2ml^2$, so that we neglect the occupation of all states except the lowest pair, denoted as $\Psi_\uparrow$ and $\Psi_\downarrow$. These two spin states encode the two logical states of the qubit.

The coupling of the spin to electric fields, central to this work, is possible through the spin-orbit interactions. 
We consider the leading, linear in momentum, terms
\begin{equation}
H_{so} = \frac{\hbar}{2m^* l}  \left( \frac{\sigma_y p_x - \sigma_x p_y}{l_r}  + \frac{- \sigma_x p_x + \sigma_y p_y}{l_d} \right),
\label{eq:so}
\end{equation}
with $l_d$ and $l_r$ the spin-orbit lengths of Rashba and Dresselhaus spin-orbit interactions, respectively. We shall parametrize these by an overall scale $l_{so}$ and a mixing angle $\nu \in [0,2\pi]$, defining $l_r^{-1} = l_{so}^{-1} \cos \nu$ and $ l_d^{-1} = l_{so}^{-1} \sin \nu$. 

If the spin-orbit interactions are weak, $l_{so} \gg l$, which is the case in GaAs and Si, the eigenstates depart only slightly from orthogonal spinor pairs, so that the state spin labels $\sigma$ are still well defined. Nevertheless, the spin-orbit separability is no longer exact and one of the consequences is the appearance of a finite dipole moment between pairs, including the lowest one:
\begin{equation}
{\bf d} = \langle\Psi_\uparrow  |  \mathbf r | \Psi_\downarrow \rangle.
\label{eq:d def}
\end{equation} 
It is straightforward (see Appendix~\ref{app:eq5deriv}) to derive the following expression for the quantum dot modeled by Eqs.~\eqref{eq:H}--\eqref{eq:so}: 

\begin{equation}
{\bf d} = \frac{ g^* \mu_B  m l^4 B}{4\hbar^2 l_{so}} \bf{v}, 
\label{eq:d exp}
\end{equation}
where $\bf v$ is the dimensionless vector  
\begin{equation}
\mathbf{v} =   \sum_{i=x',y'}\frac{l_i^4}{l^4}\left( \mathbf{n}^{i}_{so} \times \bf{b} \right)_z \mathbf{e}_{i}.
\label{eq:v def}
\end{equation}
 Here the constant vectors $\mathbf{n}^i_{so}$ are defined by writing the spin-orbit vector $\mathbf{n}_{so} = (x \sin\nu + y \cos\nu, -x\cos \nu-y\sin \nu)$ in dot coordinates $\mathbf{n}_{so} = x'\mathbf{n}^{x'}_{so}+y'\mathbf{n}^{y'}_{so}$. Furthermore, the subscript $z$ denotes the vector component perpendicular to the dot plane and $\mathbf{e}_i$ are the unit vectors of the major axis $x'$, and the minor axis $y'$ of the dot potential.

\section{Measure of qubit quality}

We now analyze how the performance of a spin qubit depends on the orientation of its  laterally gated host dot with respect to the crystallographic axes. Because of the presence of the spin-orbit interactions, the qubit described in the previous section can be controlled and manipulated electrically. This was realized, e.g., in the experiment of Ref.~\citenum{nowack2007:S}, where coherent manipulation using 
EDSR
was demonstrated. Applying a resonant electric field, the frequency of the resulting Rabi oscillations $\Omega$ quantifies the speed of the single qubit operations. However, apart from allowing for operations by local electrical fields, the presence of the spin-orbit interaction also makes the spin prone to  charge noise. The most prominent consequence is the spin lifetime $T$ being limited by coupling to the electric field of phonons, 
which dominate other mechanisms.
One can then take the pros and cons together, for example, by evaluating the number of operations that can be done within the qubit lifetime, 
\begin{equation}
\zeta  =  \Omega T.
\label{eq:FOM1}
\end{equation}
This number then allows for comparison of qubits across different samples and platforms. 

Having written the previous simple expression, we have to note that the dynamics of a qubit in the semiconductor environment is  complicated, and so is the assessment of its quality. This can be cast as the time $T$ being dependent on the task aimed at.\cite{fabian2007:APS} For example, to store classical information, it is appropriate to consider the mentioned qubit lifetime $T_1$, whereas for storage of quantum information, one should take the decoherence time $T_2$. The latter can be surprisingly long compared to inhomogeneous dephasing times $T_2^*$ \cite{merkulov2002:PRB,khaetskii2002:PRL}, if the qubit is protected from quasi-static noise by spin-echo techniques \cite{bluhm2011:NP}. 
To reflect these numerous variants, we separate the decay rate entering Eq.~\eqref{eq:FOM1} as being due to coupling to dipolar noise, and additional decay channels 
 parameterized by the rate $\Gamma_o$.  The additional decay channels impose a limit on the degree of spin relaxation that can be suppressed by the applied magnetic field. One such limitation is due to spatially varying spin-orbit couplings, resulting from, e.g.,  varying concentration of dopants \cite{sherman2005:PRB}.  Assuming weak coupling,
the two decay rates contribute in parallel,
\begin{equation}
T^{-1} = \Gamma_{\text{ph}}+\Gamma_o. 
\label{eq:T}
\end{equation} 
Which channels are to be included depends on what specific feature of the qubit is assessed by Eq.~\eqref{eq:FOM1}.

Separation of the environment induced decay into the two parts as in Eq.~\eqref{eq:T} allows us to analyze the qubit quality by studying the behavior of its dipole moment, Eq.~\eqref{eq:d def}. Namely, because of the small spatial extent of the quantum dot eigenstates (with $l$ typically tens of nanometers), the resonant electric field ${\bf E}$ induced Rabi frequency is given by
\begin{equation}
\Omega \propto | {\bf E} \cdot {\bf v} |.
\label{eq:omega}
\end{equation}
We will assume ${\bf E} || {\bf d}$, which maximizes the Rabi frequency, since a misalignment of these two vectors leads to a trivial suppression factor of the Rabi frequency. The phonon induced relaxation is
\begin{equation}
\Gamma_{\text{ph}} \approx \gamma |{\bf v}|^2,
\label{eq:gamma}
\end{equation}
with $\gamma$ a function independent on ${\bf v}$. The validity of this approximation, crucial for our further analysis, is demonstrated in detail in Appendixes~\ref{app:eq9gaasderiv} and \ref{app:eq9sideriv}. There we show that for quantum dot parameters typical for current experiments, Eq.~\eqref{eq:gamma} is an excellent approximation for both GaAs and Si/SiGe lateral quantum dots, for magnetic fields up to several tesla. On the other hand, we stress that the result in Eq.~\eqref{eq:gamma}, which substantially simplifies the optimization problem, is non-trivial. Indeed, beyond the dipole approximation, the anisotropic constants of piezoelectric phonons will result in additional directional variations of the rate, and thus the figure of merit. See Appendix~B for the discussion on the symmetry requirements that lead to Eq.~\eqref{eq:gamma}.

Equations \eqref{eq:FOM1}--\eqref{eq:gamma} result in the following figure of merit describing the qubit quality:
\begin{equation}
\zeta = \zeta_0  \frac{|{\bf v}|}{|{\bf v}|^2 + \Gamma_o/\gamma},
\label{eq:FOM2}
\end{equation} 
where $\zeta_0$ is a constant. Our goal henceforth is to find conditions on the geometry (that is, the orientation of the dot aexs and the magnetic field direction with respect to the crystallographic axes) that maximize the figure of merit. Accordingly, we refer to parameters that yield the global maximum of $\zeta$ as optimal parameters, and the set of optimal parameters as the optimal configuration.
We will also compare dots with different ellipticities and spin-orbit interactions by comparing the maximal achievable value of $\zeta$. Finally, we will assess the stability of the working point (the geometry corresponding to a maximal $\zeta$), as the sensitivity of $\zeta$ to small fluctuations of its parameters.

In these investigations, the parameters of the dot orientation ($\delta$), magnetic field orientation ($\beta$), the spin-orbit mixing angle ($\nu$), and the dot ellipticity ($\epsilon$) enter only the dipole moment. They do not enter the overall constant  $\zeta_0$ (which is then irrelevant for the figure of merit maximization) or $\Gamma_o/\gamma$.
The latter, in general a complicated function of the magnetic field strength, the spin-orbit scale, and the dot confinement, can be thus taken as a single ``external'' parameter, being the ratio of the electric dipolar to other sources of noise. With Eq.~\eqref{eq:FOM2}, we have thus reduced the number of free parameters in the qubit optimization to a reasonably small set for the problem to be tractable.

\section{Optimal dot geometry}
\label{sec:optimalgeometry}
Here we investigate the impact of the dot geometry on the figure of merit. Specifically, we wish to investigate the following questions. Is it possible to improve the figure of merit by varying the magnetic field orientation?
In the case of elliptical quantum dots, is there a dot orientation, with respect to the crystallographic axes, which is more suitable for electric control of the spin qubit?
And lastly, are there cases when the figure of merit for elliptical dots can be maximized by reorienting only the magnetic field? 
These questions are important for the dot design, as the dot orientation can not be varied once the structure is fabricated.
\begin{figure}[tp]
     \includegraphics[width=0.4\textwidth]{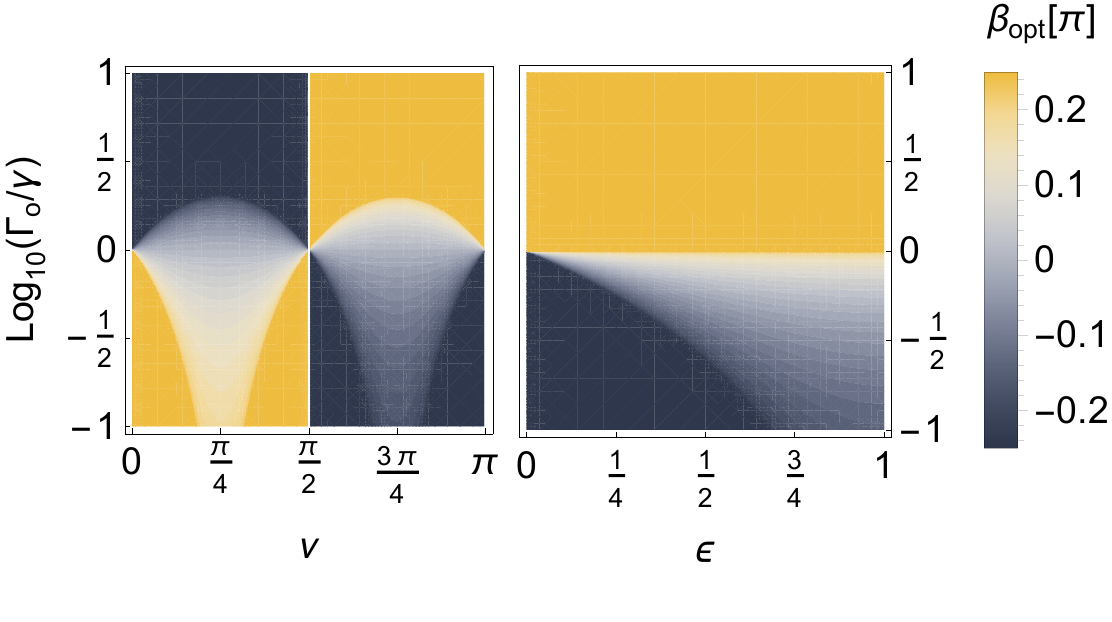}
    \caption{ (Left) Optimal orientation of the magnetic field $\beta_{opt} \in [ -\pi/4,\pi/4 ]$
 as a function of the ratio $\Gamma_0/\gamma$ and spin-orbit mixing angle $\nu$ in a circular dot ($\epsilon = 0$). 
 (Right) $\beta_{opt} $ as a function of  $\Gamma_0/\gamma$ and ellipticity $\epsilon$ for a dot with $\delta=\pi/4$ and $\nu=0$ (only the Rashba interaction is present).
}\label{fig:optimalcircular}
\end{figure}
\subsection{Circular dot}
For a circular dot, 
the figure of merit does not depend on $\delta$, which in this case means only a coordinates choice.
The angle $\beta$ of the magnetic field direction has a unique optimal value $\beta_{opt}$ that depends on the spin-orbit mixing angle $\nu$ and $\Gamma_0/\gamma$.
Specifically, by 
solving $\partial_\beta \zeta=0$, 
we obtain for $\nu \not = 0, \pi/2$, 
\begin{equation}
\beta_{opt}  =
\left \{ \begin{array}{ll}
\text{sgn}(l_r l_d)\frac{\pi}{4},  & \text{if   }   \frac{\Gamma_0}{\gamma}> 1+|\sin 2\nu|, \\
\text{sgn}(l_r l_d)\frac{3\pi}{4}, & \text{if   } \frac{\Gamma_0}{\gamma}< 1-|\sin 2\nu|, \\
\frac{1}{2}\arcsin \left(\frac{\Gamma_0/\gamma - 1 }{\sin 2 \nu} \right), & \text{otherwise}.\\
\end{array}
\right. 
\label{eq:optimalbetacirc}
\end{equation} 
For the two exceptions  $\nu = 0$ and $\nu = \pi/2$, corresponding to, respectively, the case of dominating Rashba and dominating Dresselhaus spin-orbit interaction, the figure of merit for circular dots is independent of the magnetic field orientation.

The results in Eq.~\eqref{eq:optimalbetacirc} show that by reorienting the magnetic field, it is possible to 
improve the qubit quality.
However, the optimal orientation depends on $\Gamma_0/\gamma$. 
When either $\Gamma_0$ or $\gamma$ are dominant, the figure of merit is maximized by the magnetic field direction $\beta_{opt} = \text{sgn}(l_r l_d)\pi/4$ and $\beta_{opt} = \text{sgn}(l_r l_d)3\pi/4$, respectively, which depends only on the relative sign of the spin-orbit strengths, see the left panel of Fig.~\ref{fig:optimalcircular}.
On the other hand, when the two decay channels are comparable, quantified by the condition $|\Gamma_0-\gamma|<\gamma |\sin 2\nu|$, the optimal orientation $\beta_{opt}$ requires to know both $\Gamma_0/\gamma$ and $l_r/l_d$. To conclude the case of circular dots, the figure of merit can be tuned by reorienting the magnetic field unless one of the spin-orbit interactions is absent.

\subsection{Elliptical dot}
For the more general case of elliptical dots ($\epsilon>0$), we have to take the orientation of the quantum dot, $\delta$, into account.
By differentiating the figure of merit $\zeta$ with respect to $\delta$ and $\beta$, we find that as for the circular dots the solution maximizing the figure of merit depends on the ratio $\Gamma_{0}/\gamma$ and can be separated into the three scenarios. Dominant $\Gamma_0$, dominant $\gamma$, and when $\Gamma_0,\gamma$ are comparable. We can express the optimal configuration as the parameters $\delta$ and $\beta$, which result in the dimensionless vector:
\begin{equation}
\label{eq:solutions}
|\mathbf v| =\left \{ \begin{array}{ll}
 \sqrt{1+|\sin2\nu|},  & \text{if  } \frac{\Gamma_0}{\gamma}> 1+|\sin 2\nu|,	\\
 \sqrt{ 1-|\sin2\nu| } \bar \epsilon, & \text{if  } \frac{\Gamma_0}{\gamma}< (1-|\sin 2\nu|) \bar \epsilon^2,\\
 \sqrt{\frac{\Gamma_0}{\gamma}}, & 
\text{otherwise},
\end{array}
\right.
\end{equation}
where $\bar \epsilon = 1-\epsilon$. From the first two solutions, we immediately find a unique optimal configuration, depending only on the sign of the spin-orbit lengths. Explicitly, the optimal configurations are $\delta_{opt} = \text{sgn}(l_r l_d)\pi/4 $ and  $ \beta_{opt} =  \pm \text{sgn}(l_r l_d)\pi/4$ for the first and second cases, respectively.
These optimal configurations are unique and therefore a specific
quantum dot orientation is required to maximize the figure of merit.

\begin{table} 
\begin{tabular}{c | r| r}
\hline
$l_r/l_d$&
$\delta_{opt}$ &
$\beta_{opt}$ 
 \\ \hline
$\gg$1    	&   $ arb. $	    & $-\delta\pm \frac{1}{2}\text{Re}[\arccos (-\xi_1)]$    \\
$\ll$1 				&	$ arb. $		& $\delta \pm \frac{1}{2} \text{Re}[\arccos \xi_1]$  	 \\
$ \pm 1 $    & 	$ \pm \pi/4 $	& $\pm \frac{1}{2} \text{Re} [\arcsin \xi_2]$  	 \\
\hline
\end{tabular}
\caption
{ Parameters for optimal layout of an elliptical quantum dot in the cases of dominating Rashba spin-orbit interaction, dominating Dresselhaus and equal magnitude of the two. Here Re denotes the real part, $\xi_2 = \Gamma_0/\gamma -1$  and $\xi_1 = 1 + 2\xi_2/(1-\bar \epsilon^2)$ with $\bar \epsilon = 1 -  \epsilon$.
}
\label{tab:optimaltable}
\end{table}

Conversely, when $\Gamma_0 \approx \gamma$ where the third solution in Eq. (\ref{eq:solutions}) applies, there exists a range of configurations, which maximize the figure of merit.
To find an optimal configuration across all parameters, we first consider the quantum dot orientation $\delta_{opt} = \text{sgn}(l_r l_d) \pi/4$, in line with the previous two solutions. The optimal orientation of the magnetic field is then $\beta_{opt} = -\delta_{opt} \pm \chi /2$, where
\begin{equation}
\label{eq:ellipticalchi}
 \chi = \arccos \left ( \frac{\frac{\Gamma_0}{\gamma} - \sin^2( \tilde  \nu  +\frac{\pi}{4})- \bar \epsilon^2 \sin^2 (\tilde  \nu-\frac{\pi}{4})}{\bar \epsilon^2 \sin^2 (\tilde  \nu - \frac{\pi}{4}) - \sin^2 (\tilde  \nu + \frac{\pi}{4})} \right),
\end{equation}
with $\tilde  \nu = \text{sgn}(l_r l_d) \nu$.
This implies that with a dot orientation satisfying $\delta = \text{sgn}(l_r l_d)\pi/4$ we can obtain the global maximum of the figure of merit by reorienting the magnetic field irrespective of the value of $\Gamma_0/\gamma$. 
This result shows that there are quantum dot orientations more suitable for electric control of the spin qubit, which answers the second question.

Since it is usually difficult to extract the values of spin-orbit lengths precisely  in a quantum dot \cite{raith2014:PSSB,scarlino2014:PRL}, we now consider three limiting cases
 of, respectively, dominant Rashba, dominant Dresselhaus, and equal spin-orbit strengths. The optimal parameters for these cases, are given in Table \ref{tab:optimaltable} and can be used as general guidelines for maximizing the figure of merit when only a rough estimate of the spin-orbit strengths is available.  

Let us start with the case of an equally strong Dresselhaus and Rashba spin-orbit coupling, $l_r = \pm l_d$. Here the figure of merit can be maximized for a range of quantum dot orientations $\delta$, where the range depends on the ellipticity parameter $\epsilon$ and $\Gamma_0/\gamma$ ratio. Interestingly, the dot orientation $\delta_{opt}=\text{sgn}(l_r l_d) \pi/4$ ensures that it is always possible to reach an optimal configuration which does not depend on the dot ellipticity.

On the other hand, if Rashba or Dresselhaus spin-orbit interaction is dominant, the global maximum of the figure of merit can be reached regardless of the dot orientation. This can be seen by inspecting the angle dependence of the dipole moment $\mathbf d$, which is contained in the z component of the cross product $\mathbf n_{so}(\delta,\nu) \times \mathbf B(\beta)$. Under the condition that 
$R_z(\Lambda)\mathbf n_{so}(\delta,\nu) = \mathbf n_{so}(0,\nu)$,
where $R_z(\Lambda)$ denotes rotation about the z axis by angle $\Lambda$, the equality  
\begin{equation*}
\left[R_z(\Lambda) \mathbf n_{so}(0,\nu) \times \mathbf B(\beta)\right]_z  =\left[\mathbf n_{so}(0,\nu) 
\times \mathbf B(\beta-\Lambda)\right]_z
\end{equation*} 
then ensures that $\mathbf d$ is fully controllable by the magnetic field direction.
This condition is fulfilled only for $\nu = 0$ and $\pi/2$ (with $\Lambda = \delta$), where the resulting dipole moment $\mathbf d$ depends only on the sum or difference of the magnetic field and quantum dot orientation angles, respectively.
Since the parameter dependence of the figure of merit is entirely contained in $\mathbf d = \mathbf d(\delta ,\beta)$, it follows that $\zeta(\delta,\beta) = \zeta(\beta - \delta)$. 
This implies that the figure of merit for a quantum dot with an arbitrary orientation can always be maximized, and the optimal configuration can be achieved, by only reorienting the applied magnetic field if one of the spin-orbit interactions is dominant.
This answers the third question. 

\begin{figure}
\includegraphics[width=0.45\textwidth]{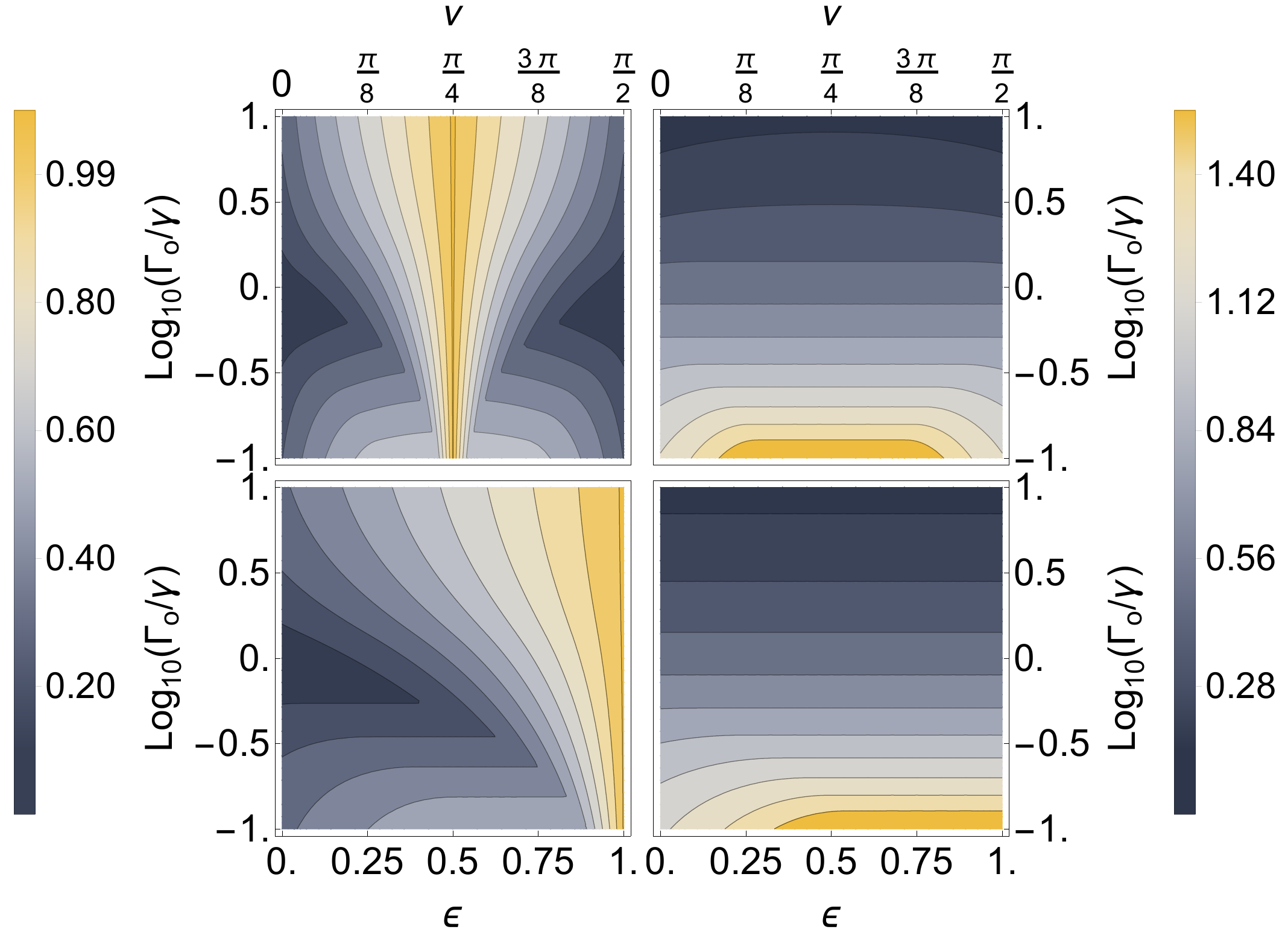}
  \caption{The tunability $\mathcal{T}$ (left column) and maximal quality  $\mathcal{Q}$  (right column). With $\epsilon = 0.25$  (top) and spin length ratio $l_r/l_d = 5, 1/5$ (bottom). } 
  \label{fig:additionaloptimization}
\end{figure}

\begin{figure}[tp]
\includegraphics[width=0.48\textwidth]{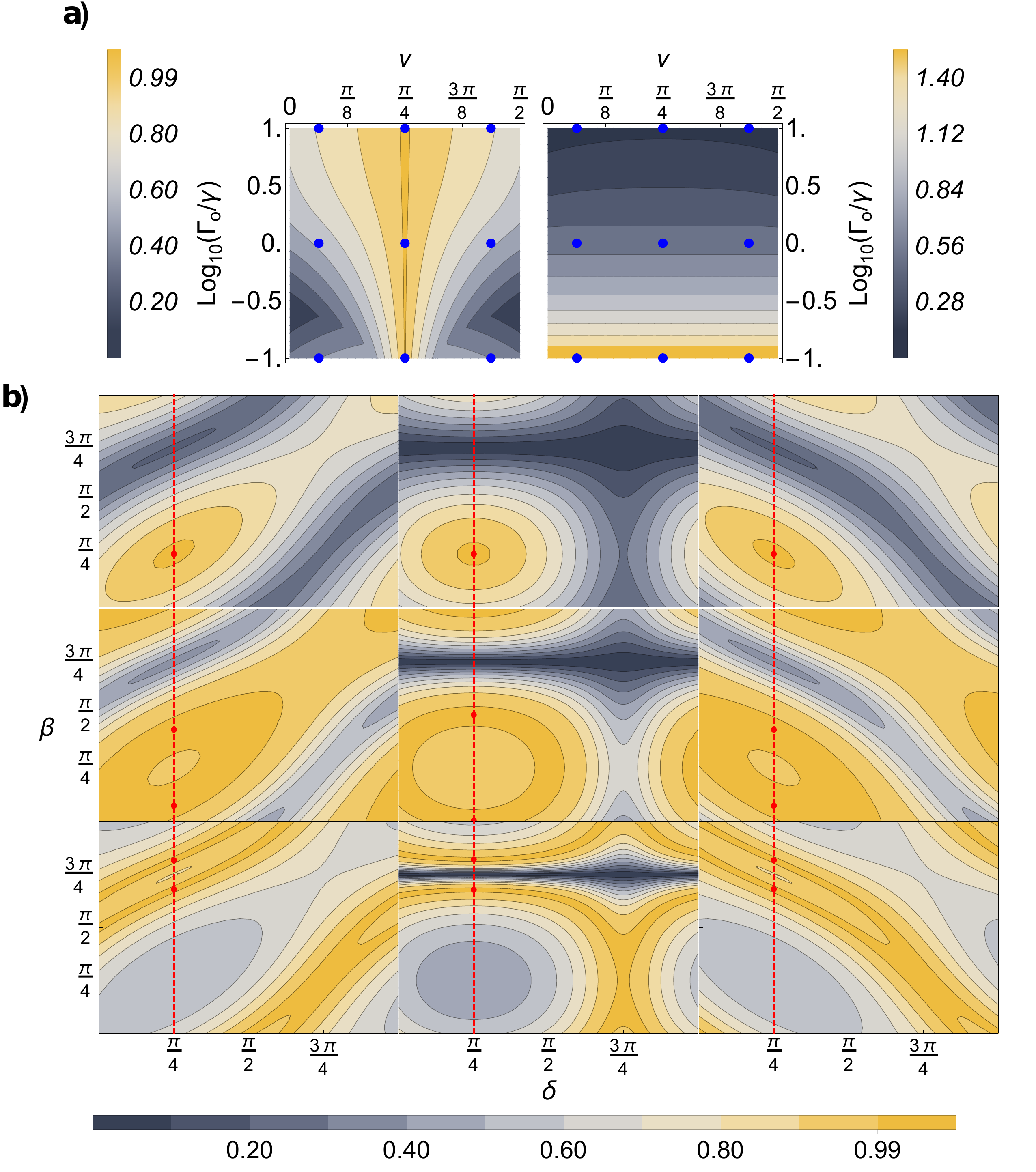}
\caption{ (a) Left figure shows the tunability of the figure of merit, $\mathcal{T} $. Right figure shows the maximal quality  $\mathcal{Q}$. Each blue point corresponds to the parameters ($\nu, \Gamma_0/\gamma$) of the corresponding figure in (b).
(b) The normalized figure of merit $\zeta=\zeta(\delta,\beta)$ is displayed in a $3 \times 3$ array of figures. In all figures elliptic quantum dots with $\epsilon = 0.75$ are considered in host materials where the Rashba and Dresselhaus spin-orbit couplings have the same sign, i.e., $l_r l_d > 0$. 
The red dashed vertical line shows an optimal dot orientation with red points indicating the optimal configuration ($\delta_{opt}, \beta_{opt}$).}
\label{fig:optimization}
\end{figure}

\subsection{Discussion}

We now discuss our results in more detail with the help of Figs. \ref{fig:additionaloptimization}  and \ref{fig:optimization}.
The first to note is that the figure of merit $\zeta$ strongly depends on the quantum dot geometry and the ratio of spin-orbit lengths.
With the results for circular dots in Eq.~\eqref{eq:optimalbetacirc} we already know that the gain from the optimization of $\beta$ and $\delta$ varies for different dots. For example, for a circular dot with dominant Rashba or Dresselhaus spin-orbit interaction, the figure of merit is independent of these parameters.
An important question is then, when is the potential gain of our optimization scheme large? 

We address this question by evaluating tunability $\mathcal{T}$ and maximal quality $\mathcal{Q}$.
The former is defined as $\mathcal{T} = (\max_{\beta, \delta}\zeta-\min_{\beta, \delta}\zeta)/\max_{\beta, \delta}\zeta \in [0,1]$ and relates the maximum and minimum of the figure of merit as a function of $\beta$ and $\delta$. For $\mathcal{T}\ll 1$, $\zeta$ does not change much when varying $\delta$ and $\beta$, and accordingly there is little to gain from varying these parameters. On the other hand, for $\mathcal{T}\approx 1$, the dot is highly tunable, with $\zeta$ changing significantly upon varying $\beta$ and $\delta$. The tunability is plotted
in the left column of Fig.~\ref{fig:additionaloptimization}. Here we see that it increases with increasing dot ellipticity, and is maximal when the Rashba and Dresselhaus spin-orbit lengths are the same. 

To compare optimized qubits across different parameters, we define the maximal quality of the dot as $\mathcal{Q} = \max_{\beta, \delta}\zeta/\zeta_0$. Qubits with larger $\mathcal{Q}$ are better (we assume the value of $\gamma$ is the same for the two qubits being compared). As seen in
the right column of Fig.~\ref{fig:additionaloptimization}, $\mathcal{Q}$ increases if the phonon-induced relaxation rate scale $\gamma$ is larger than the rate of additional decay channels $\Gamma_o$. To understand this, it is useful to consider the limit where the additional channels are completely absent. The Rabi frequency and the phonon induced relaxation rate are then proportional, respectively, to the first and second powers of the qubit dipole moment $|{\bf v}|$. By tuning $|{\bf v}|$ to zero, we can make their ratio arbitrarily large. Decreasing ${\bf v}$ thus improves the qubit quality until the phonon induced relaxation drops below other decay channels. In addition to this, the figure also shows that the previous two conditions which increase tunability, namely equal spin-orbit lengths and high ellipticity, also increase the maximal quality.

Having established when the prospective gains are large, the next question is how to design the quantum dot structure in order to maximize $\zeta=\zeta(\delta,\beta)$ using the two parameters at our disposal.
We show a representative case, an elliptic quantum dot with ellipticity $\epsilon= 0.75$, in Fig.~\ref{fig:optimization}. In Fig.~\ref{fig:optimization}(a) the corresponding tunability (left) and maximal quality (right) are displayed. 
For each blue point in panel (a), the figure of merit is plotted as a function of $\delta$ and $\beta$,
resulting in the array of figures in Fig.~\ref{fig:optimization}(b). 
There we see that
 the location and spreading of the optimal configuration in the parameters space $(\delta, \beta )$ depends in general on the ratio $\Gamma_{o}/\gamma$ and spin-orbit mixing angle $\nu$. 

The case of dominating phonon-induced spin relaxation, i.e., $\Gamma_{o}\ll \gamma$, displayed in the top row of Fig.~\ref{fig:optimization}(b), is characterized by a unique optimal configuration point $(\delta_{opt},\beta_{opt})$.
This indicates that a unique quantum dot orientation is necessary to maximize the figure of merit.
In contrast, in the second and third rows of figures (i.e., the cases when $\Gamma_{o} \approx \gamma $ and $\Gamma_{o} \ll \gamma$) the optimal configuration is no longer unique. In these two scenarios, two new features appear. The first is that for each dot orientation there are two optimal magnetic fields $\beta_{opt}$. The second feature is that the optimal orientation of the dot, necessary to maximize the figure of merit, starts to depend on the spin-orbit lengths. We elaborate on this by examining the role of the spin-orbit mixing angle.

For equal spin-orbit lengths, displayed in the center column of Fig.~\ref{fig:optimization}(b), 
there exists a magnetic field orientation $\beta_{dec} = \text{sgn}(l_r l_d)3\pi/4$, which decouples the spin from the electric field for any value of $\Gamma_0/\gamma$. 
This is the well-known symmetry beyond the persistent spin helix \cite{schliemann2003:PRL,bernevig2006:PRL,duckheim2007:PRB,koralek2009:N,duckheim2009:PRB}.
In this case, the magnetic field should be oriented perpendicular to this decoupling direction.
Moreover, the quantum dot orientation $\delta = $sgn$(l_r l_d)3\pi/4$ results in a vanishing figure of merit for $\epsilon= 1$ [not shown in Fig. 4(b)],  irrespective of magnetic field orientations. It is so because there is no spin-orbit field for momentum along the direction set by such $\delta$.

In the opposite case, for dominant Rashba spin-orbit interaction, shown for specific $\delta$ in the right panel of Fig.~\ref{fig:optimalcircular} and for specific $\epsilon$ in the left column of Fig.~\ref{fig:optimization}(b) (the case of a dominant Dresselhaus term, shown in the right column, is related by symmetry and therefore we do not discuss it separately), the orientation of the quantum dot can be compensated for, as discussed in Sec. \ref{sec:optimalgeometry} B.
This effect can be seen in the bottom left figure. Namely, as the spin-orbit length ratio approaches the asymptotic values $l_r/l_d \rightarrow  \infty$ and $l_r/l_d \rightarrow  0$, the optimal configurations start to lie along diagonal stripes in the $(\delta, \beta)$ parameter space.
 Consequently, in this scenario, the figure of merit can always be maximized by reorienting the magnetic field alone.

To summarize the case of elliptic dots, we have found that the orientations of the magnetic field and quantum dot play a crucial role for highly elliptical dots $(\epsilon \approx 1)$ with dominant phonon-induced spin relaxation $(\Gamma_0/\gamma \ll 1)$ and with comparable spin-orbit lengths $(\nu \approx \pi/4)$. In this case, the potential gain from the optimization is the largest. 
We also note that dominant Rashba or Dresselhaus spin-orbit interaction permits a wider range of quantum dot orientations where an optimal configuration is possible, but at the cost of a smaller gain.

We conclude this discussion by highlighting two robust features, displayed also in Fig.~\ref{fig:optimization}b. Despite the different behavior of $\zeta$ as we vary parameters, there exists an overall optimal quantum dot orientation $\delta_{opt} = \text{sgn}(l_r l_d) \pi/4$. This orientation guarantees that $\zeta$ can be maximized by varying only the magnetic field direction. 
For a quantum dot with a magnetic field tunable in the quantum dot plane, $\delta_{opt}$ provides the preferred major axis of an elliptic dot (or the inter-dot vector in the case of a double dot). 
Second, for the parameters of Fig.~\ref{fig:optimization} (moderately elliptic dots), the optimal magnetic field orientation depends mostly only on $\Gamma_o/\gamma$. One can therefore pick a (nearly) optimal configuration from knowing only the value of $\Gamma_o/\gamma$ and the relative sign of the spin-orbit interactions.

\section{Conclusions}

We have investigated the possibility of optimizing EDSR based electrical control of spin qubits for quantum dots in GaAs and Si/SiGe by exploiting anisotropy of the spin-orbit interactions.
We asses the quality of the qubit by a figure of merit, being the ratio between the Rabi frequency and the spin relaxation rate. In the dipole limit, which is justified for typical gated GaAs and Si/SiGe quantum dots, the figure of merit can be entirely parameterized by the ratio of Rashba and Dresselhaus spin-orbit couplings, the quantum dot ellipticity, and the overall spin relaxation rate magnitude.

We have found that the ratio of the two dominant spin-orbit couplings has the most significant impact on the optimal configuration. In particular, for the case of dominant Rashba or Dresselhaus spin-orbit interaction,  the dot can be optimized by reorienting the applied in-plane magnetic field irrespective of the quantum dot orientation.
For the remaining cases, we find a unique dot orientation which ensures that the figure of merit can be maximized by changing the magnetic field orientation. This allows us to propose $\delta_{opt} = \text{sgn}(l_r l_d) \pi/4$ as an overall optimal quantum dot orientation.

Our results show that a straightforward adjustment of quantum dot geometry and orientation of the magnetic field can substantially increase the efficiency of electrical control of a spin qubit.

\acknowledgments
We would like to thank L. M. K. Vandersypen for inspiring this work and for valuable comments on the manuscript.
O. Malkoc acknowledges support from JSPS Postdoctoral Fellowship for Overseas Researchers. 
This work is supported by the JSPS Kakenhi Grant No. 16K05411, Swiss NF, NCCR QSIT, and IARPA..

\appendix
\section{Derivation of Eq. (\ref{eq:d exp})}
\label{app:eq5deriv}
Here we derive the expression for the dipole moment, Eq.~\eqref{eq:d exp}.
 Since the spin-orbit interaction is weak in both GaAs and Silicon we treat the interaction perturbatively.
Starting without spin-orbit interaction, the orbital and spin parts of the wavefunction are separable.  
With the bi-harmonic quantum dot confinement potential given in Eq.~\eqref{eq:potential} the Hamiltonian acting on the orbital part is described by two harmonic oscillators with characteristic frequencies $\omega_{x'} = \hbar/m l_{x'}^2,  \omega_{y'} = \hbar/m l_{y'}^2$, where $l_{x'} = l$ and $l_{y'} = l(1-\epsilon)^{1/4}$. The in-plane magnetic field, $\mathbf B = B\mathbf b$, induces a Zeeman splitting and acts only on the spin state. 
The Hamiltonian without spin-orbit interactions is
\begin{equation}
H_{0}
 =\hbar \omega_{x'} \left (  n_{x'} + \frac{1}{2} \right ) + \hbar \omega_{y'} \left (  n_{y'} + \frac{1}{2} \right )
+ \frac{g^*\mu_B  {\bf B}\cdot \boldsymbol{\sigma} }{2},
\end{equation}
where $ n_{x'} =  a^\dagger_{x'}  a_{x'}$, $ n_{y'} =   a^\dagger_{y'}  a_{y'}$ are number operators of the two harmonic oscillators. This Hamiltonian is diagonal in the basis of 
$|\phi_{n_{x'} n_{y'} s}\rangle = |n_{x'} n_{y'}\rangle | s\rangle $, the harmonic oscillator eigenstates with a definite spin projection $\sigma=\pm 1$ along the magnetic field ${\bf B}$.
To obtain a Hamiltonian with spin-orbit interactions which is more suitable for the perturbative expansion \cite{baruffa2010:PRL}, we perform a Schrieffer-Wolff transformation with the unitary operator 
  $ U=\exp(i \mathbf{{n}}_{so} \cdot {\boldsymbol \sigma}/2 )$, where
$\mathbf{{n}}_{so} = ( x l^{-1}_d +  yl^{-1}_r, - xl^{-1}_r - yl^{-1}_d,0)$, and transform $ H={H}_0+{H}_{so}$ into an effective Hamiltonian 
\begin{equation}
{H} \rightarrow {U} {H} {U}^{\dagger} = {H}_{eff}.
\end{equation}
The transformed Hamiltonian is the same as ${H}$, with the exception of the spin-orbit term ${H}_{so}$ which is replaced by the effective interaction term \cite{baruffa2010:PRL}
\begin{equation}
{H}^{\text{so}}_{\text{eff}} =  \frac{g^* \mu_B}{2}(\mathbf{{n}}_{so} \times \mathbf{B} ) \cdot \boldsymbol{ {\sigma}}  - \frac{\hbar^2}{4m}  \left[  \frac{1+l_z {\sigma}_z }{l_r^2}  + \frac{1-l_z {\sigma}_z}{l_d^2} \right ].
\end{equation}
Here $ l_z  = -i( x \partial_y -  y \partial_x)$ is the orbital momentum in units of $\hbar$. 
The first-order correction to the unperturbed wave functions is then
\begin{equation}
\label{eq:wfcorr}
 \delta| \Psi_{\sigma} \rangle  =  U  \sum \limits_{\alpha \not = 00\sigma} \frac{\langle \phi_{\alpha} |{H}^{so}_{eff}|\phi_{00\sigma}\rangle}{\epsilon_{00 \sigma} - \epsilon_{\alpha}}|\phi_{\alpha}\rangle,
\end{equation}
where we have introduced the composite index $\alpha = \{n_{x'}, n_{y'}, s\}$. Using Eq.~\eqref{eq:wfcorr} with the definition of the dipole moment in Eq.~\eqref{eq:d def} we obtain  
\begin{equation}
\label{eq:dipole}
\mathbf d = \frac{g^* \mu_B  m l^4 B}{4\hbar^2 l_{so}}  \sum_{i=x',y'}\frac{l_i^4}{l^4}\left( {\bf n}^{i}_{so} \times {\bf b} \right)_z \mathbf{e}_{i}   + \mathcal{O}\left(\frac{\varepsilon_z}{\hbar \omega_i}\right),
\end{equation}
where $\mathbf{e}_{x'} (\mathbf{e}_{y'})$ is the unit vector along the major (minor) axis of the dot, $\varepsilon_z=g^* \mu_B B $ is the Zeeman energy and ${\mathbf{n}}_{so} = {x}'\mathbf{n}_{so}^{x'}+{y}'\mathbf{n}_{so}^{y'}  $. In our work we restrict our analysis to the limit where the orbital excitation energies are much greater than the Zeeman splitting, $\hbar \omega_{x'/y'}\gg  \varepsilon_z$.
In particular, for quantum dots considered in this work, with typical confinement energy $\sim 1$ meV, the higher-order corrections in Eq.~\eqref{eq:dipole} are small and we neglect them. The dipole moment is then,
\begin{equation}
{\bf d} =\frac{g^* \mu_B  m l^4 B}{4\hbar^2 l_{so}} \mathbf v, 
\end{equation}
where $\mathbf v$ is the dimensionless vector given in Eq.~\eqref{eq:v def}.

\section{Derivation of Eq. (\ref{eq:gamma}) for GaAs}
\label{app:eq9gaasderiv}

To calculate the spin relaxation rate, we follow the standard procedure suitable for zero temperature \footnote{A small finite temperature will contribute a multiplicative (Boltzmann) factor to the rate which does not change its directional anisotropy. At higher temperatures, two-phonon processes might become relevant.\cite{kornich2014:PRB}} and Zeeman energies corresponding to magnetic field not larger then a few tesla \cite{cheng2004:PRB,sherman2005:PRB,prada2008:PRB,meza-montes2008:PRB,wang2010:PRB}.
In this regime, the dominant phonon induced spin relaxation is due to acoustic phonons. The coupling between the longitudinal acoustic and transversal acoustic phonons of the host crystal and the confined electrons is
\begin{equation}
\label{eq:phononcoupling}
{H}_{\text{ph}} = i \sum \limits_{\mathbf K \lambda} \sqrt{\frac{\hbar K}{2 \rho V c_\lambda}}  
M^\lambda_{\mathbf K } \left[   {b}^\dagger_{\mathbf K,\lambda} e^{i \mathbf K \cdot {\mathbf{R}}} - {b}_{\mathbf K,\lambda} e^{-i \mathbf K \cdot  {\mathbf{R}}}\right],
\end{equation}
 where $\lambda = l, t_1, t_2$ denotes the longitudinal and two transversal polarizations of the phonons, $\mathbf K$ is the phonon wave vector, ${\mathbf{R}} = ( x, y, z)$ is  the position vector in the crystallographic coordinates, $ b$ is the phonon annihilation operator, $c_\lambda$ is the speed of sound, $\rho$ is the mass density, $V$ is the crystal volume, and $M^\lambda_{\mathbf K }$ is the material dependent piezoelectric and deformation potential.
We evaluate the spin relaxation rate by the Fermi's golden rule,
\begin{equation}
\label{goldenrule}
\Gamma_{\text{ph}} = \frac{\pi}{\rho V} \sum \limits_{\mathbf{K} \lambda}  \frac{ K \left |M_{\mathbf{K}}^\lambda  \right |^2 }{ c_\lambda}   |\langle  \Psi_\downarrow |e^{i \mathbf K \cdot  {\mathbf{R}}} | \Psi_\uparrow \rangle |^2  \delta (\hbar \omega_{ K}^\lambda - \varepsilon_z ).
\end{equation} 
The piezoelectric and deformation potentials entering $M_{\mathbf{K}}^\lambda $ inherit the symmetries of the host material.
The phonon energies are $\hbar \omega_{ K}^\lambda = \hbar c_\lambda K $. 
We employ the dipole approximation $\exp(i \mathbf{K} \cdot {\mathbf{R}}) \approx 1 + i (\mathbf K \cdot {\mathbf{R}})$, as the wavelength of the phonons corresponding the Zeeman energy is much greater than the dot confinement length along the major axis, $l$. In the continuum limit, where  $\sum_\mathbf{k} \rightarrow V(2 \pi)^{-3}\int d^3\mathbf K$, Eq.~\eqref{goldenrule} gives the spin relaxation rate
\begin{equation}
\label{eq:phononrelax}
\Gamma_{\text{ph}} = \sum \limits_{ \lambda}
 \int d^3 \mathbf K  \frac{K}{8\pi^2 c_\lambda}  |M_{\mathbf K}^{ \lambda}|^2 
 |  \mathbf K \cdot \mathbf d |^2 \delta (\hbar \omega^{\lambda}_{K} -    \varepsilon_z ).
\end{equation}
Concerning spin relaxation, an essential property of the GaAs crystal structure is the lack of inversion symmetry.  This allows for relaxation via piezoelectrical acoustic phonons. As a consequence the geometric factor in GaAs,
\begin{equation}
M^\lambda_{\mathbf K } =
 \sigma_e \delta_{\lambda,l} - 2i e h_{14} K^{-3} (K_y K_z, K_z K_x, K_x K_y)\cdot \mathbf{e}^{\lambda},
\end{equation}
is generally anisotropic. The phonon polarization unit vectors are $\mathbf{e}^{l} = K^{-1}(K_x,K_y,K_z),\ \mathbf{e}^{t1} =\mathbf{e}_z \times \mathbf{e}_{l} /|\mathbf{e}_z \times \mathbf{e}_{l}|,\  \mathbf{e}^{t2} = \mathbf{e}_l \times \mathbf{e}_{t2} $ and $eh_{14}$ is the piezoelectric coefficient. 
Because the geometric factors enter as $|M^\lambda_{\mathbf K } |^2$ in Eq.~\eqref{eq:phononrelax}, the spin relaxation rate can be decomposed into two separate contributions, one due to the isotropic deformation potential and one due to the anisotropic piezoelectric potential.
Carrying out the integral in Eq.~\eqref{eq:phononrelax}, we see that both contributions result in a relaxation rate that can be written as a sum of two terms:
\begin{equation}
\label{additive}
\Gamma_{\text{ph}} = \Gamma_{\text{ph}}^{x'} + \Gamma_{\text{ph}}^{y'},
\end{equation}
corresponding to relaxation rates with the dipole moment $\bf d$ in Eq.~\eqref{eq:phononrelax} replaced by $d_{x'} \mathbf{e}_{x'}$, and $d_{y'} \mathbf{e}_{y'}$, respectively.
The absence of cross terms in the anisotropic piezoelectric term can be understood by considering the crystal symmetry of GaAs, described by the zinc-blende structure.
The point group of a zinc-blende structure contains the two-fold rotational symmetry $C_2$ about the crystallographic axes \citep{yu2010fundamentals}.  For our argument we consider the rotation about the $[010]$ or $y$ axis, which can be represented by the operator $A_1: (x,y,z)\rightarrow(-x,y,-z)$. 
Because of the $C_2$ symmetry, the geometric factor $|M_{\mathbf K}^\lambda|^2$ is invariant under the transformation $A_1$. Since evaluating the relaxation rate in any coordinate system produces the same result, the difference between the integrals evaluated in coordinate systems $(x,y,z)$ and $A_1(x,y,z)$ vanishes. From this condition we obtain that the mixed terms $ \int d^3 \mathbf K \sum_{ \lambda} |M_{\mathbf K}^\lambda|^2 K_{x} K_{y} d_{x} d_{y}$ vanish identically. The result in Eq.~\eqref{additive} is then obtained by also taking into account the reflection symmetry with respect to the $(1\bar 10)$ plane, $A_2:(x,y,z)\rightarrow(y,x,z)$. 
From these symmetry arguments, we 
infer that if the dipole approximation is valid, the relaxation rate is proportional to only the magnitude of the dipole moment (though we considered phonons explicitly, this result can be generalized to include any dipolar noise which obeys the crystal symmetries \cite{semenov2004:PRL,san-jose2006:PRL}):
\begin{equation}
\label{eq:GammaGaasProp}
\Gamma_{\text{GaAs}} \propto |\mathbf d|^2.
\end{equation}
In the main text, we have introduced $\gamma$ as the proportionality constant $\Gamma_{\text{GaAs}} = \gamma |\mathbf v|^2$. 
We stress that abandoning the dipole approximation, will result in correction terms that do introduce additional anisotropy.
However, since the lowest-order relative correction to the relaxation rate beyond the dipole limit is in our model
$\delta \Gamma_{\text{ph}} / \Gamma_{\text{ph}} \sim K^2 l^2$, 
for typical parameters of GaAs, cf. Table \ref{tab:parameters}, $K = \mu_B/ \hbar c_{l/t} \sim 10^7 $ $\text{m}^{-1}$ and $l = 30$ nm, we get that the correction is very small, smaller than $\sim (1/10)^2$, for magnetic fields up to order tesla. Similarly, corrections are to be expected upon a departure from the inversion symmetric confinement potential, however, their magnitude is difficult to estimate without knowing more details about the potential shape.

The robust relation in Eq.~\eqref{eq:GammaGaasProp} shows that despite an anisotropic electron-phonon interaction, the anisotropy of the relaxation rate in quantum dots is exclusively due to the spin-orbit coupling. This is a substantial simplification which allows for a tractable analysis, and relatively simple analytical results for the optimal geometry applicable to a wide variety of gated quantum dots.
We note in passing that while these results have been derived for GaAs quantum dots, we have only used a few of the zinc-blende symmetries. The results are therefore valid in other host materials, where the spin relaxation due to piezoelectric acoustic phonons is dominant, if these symmetries are present. 

\section{Derivation of Eq. \ref{eq:gamma} for Si/SiGe}
\label{app:eq9sideriv}
In contrast to GaAs, silicon possesses a crystal structure with inversion symmetry. As a result, the coupling between electrons and acoustic phonons, in Eq.~\eqref{eq:phononcoupling}, is exclusively due to the deformation potential.
For bulk silicon this results in an isotropic and, compared with the piezoelectric electron-phonon coupling of GaAs, a weaker electron-phonon coupling for small Zeeman energy.
However, in the text, we consider a Si/SiGe heterostructure grown along [001]. The interface of the heterostructure causes a lattice mismatch, inducing a bi-axial strain in the growth plane. 
The strain breaks the valley degeneracy of bulk silicon and leaves a two-fold conduction band minimum located at $\mathbf K \approx \pm 2 \pi( 0,0,0.84)/a_0$, where $a_0$ is the lattice constant \cite{yu2010fundamentals}. 

We then only have to consider the (now anisotropic) electron-phonon coupling at the band minimum, which can be parametrized by two potential coefficients, $\Xi_d$ and $\Xi_u$ \cite{yu2010fundamentals}. Here, $\Xi_d+\Xi_u$ is the volume deformation potential and $\Xi_u$ is the shear deformation due to a strain along one of the valley minima. With this parametrization the electron-phonon coupling is given by 
\begin{equation}
\label{deformation}
{H}_{def} = \Xi_{d} \text{Tr}{\varepsilon} + \Xi_{u} \boldsymbol \kappa \cdot  \varepsilon \cdot \boldsymbol \kappa.
\end{equation}
Here, $\boldsymbol \kappa$ is the unit vector parallel to $[001]$ and $\varepsilon_{ij} =  (\partial_{x_j}  u_i +\partial_{x_i}  u_j)/2$  is the strain tensor, where $u_i$ is the $i-$th component of the ion displacement vector 
\begin{equation}
{\mathbf{u}} = i \sum \limits_{\mathbf K \lambda} \sqrt{\frac{\hbar K}{2 \rho V c_\lambda}}  [   {b}^\dagger_{\mathbf K,\lambda} e^{i \mathbf K \cdot {\mathbf {R}}  } - {b}_{\mathbf K,\lambda} e^{-i \mathbf K \cdot  {\mathbf{R}}} ]\mathbf{e}^{\lambda}_{\mathbf K}.
\end{equation}
The geometric factor of the electron-phonon interaction becomes
\begin{equation}
\label{newm}
M^{\lambda}_{\mathbf K} = \Xi_d \mathbf{e}_{\mathbf K}^\lambda \cdot \mathbf{e}^l_{\mathbf{K}} + \Xi_u (\mathbf{e}^\lambda_{\mathbf K})_z (\mathbf{e}^l_{\mathbf{K}})_z .
\end{equation}
Adopting again the dipole approximation 
and using Eq.~\eqref{newm}, Eq.~\eqref{eq:phononrelax} in a crystallographic spherical coordinate system gives an integral of the form
\begin{equation}
\label{phononrelax}
\Gamma_{\text{ph}} \propto \int d \theta d \phi f(\theta, \phi)
| \mathbf{e}^l_{\mathbf{K}} \cdot \mathbf d |^2,
\end{equation}
where the $\delta$ function $\delta (\hbar \omega_{ K}^\lambda - \varepsilon_z )$ has been used to fix the phonon momentum magnitude, and $f(\theta, \phi)$ describes the angle dependence of the geometric factor. For our case of a conduction minimum located along $ \boldsymbol \kappa$, we obtain 
\begin{equation}
f(\theta, \phi) = \frac{|\Xi_d + \Xi_u (\mathbf{e}_{\mathbf K}^l)^2_z |^2 }{c^5_{l}}
+
\frac{|\Xi_u(\mathbf{e}_{\mathbf K}^l)_z (\mathbf{e}_{\mathbf K}^{t_2})_z |^2}{c^5_{t}}.
\end{equation}
Since the unit vector $\mathbf{e}_{\mathbf K}^{t_2}$ does not have a z component, 
we obtain that the second term vanishes and $f(\theta,\phi) = f(\theta)$, i.e.,  the electron-phonon coupling is isotropic in the dot plane. As such, the phonon induced spin relaxation for Si/SiGe also satisfies the proportionality
\begin{equation}
\label{SiProp}
\Gamma_{\text{ph}} \propto |\mathbf d|^2.
\end{equation} 
This result shows that 
the figure of merit obtained for GaAs quantum dots is directly applicable for Si/SiGe heterostructures. The proportionality in Eq.~\eqref{SiProp} holds for quantum dots grown along any of the three main crystallographic axes, due to the cubic crystal structure of silicon.

\section{Derivation of Eq. (\ref{eq:omega})}Electrical control of the quantum dot qubit is achieved by an oscillating electric field $\mathbf E = \mathbf {E_0} \cos (\omega t)$, 
resulting in the Rabi Hamiltonian 
\begin{equation}
 {H}_R = e \mathbf E_0 \cdot {\mathbf{r}}  \cos (\omega t).
 \end{equation} The induced Rabi oscillations have a corresponding Rabi frequency $\Omega$. 
Including only the lowest two levels, to the leading order in the effective spin-orbit interaction the Rabi frequency without detuning is 
\begin{equation}
\label{rabifreq}
|\Omega|  = \hbar^{-1}\left | e \mathbf E_0 \cdot  \langle\Psi_\uparrow  |  {\mathbf{r}} | \Psi_\downarrow \rangle\right |   = \hbar^{-1}\left | e \mathbf E_0 \cdot  \mathbf d \right |.
\end{equation} 
As explained in the text, we assume that $\mathbf E$ is parallel to $\mathbf d$, which finally gives the Rabi frequency as
\begin{equation}
\Omega = \frac{e E_0}{\hbar}|\mathbf d|,
\end{equation}
where $E_0$ is the amplitude of the electric field.

\section{Parameters of GaAs and Si/SiGe}

\begin{table}[ht]
\begin{tabular}{c | r| r | r}
\hline
Parameter &
GaAs &
Si/SiGe & Units
 \\ \hline
   $g^*$	 				   		 & -0.44   & 2 &  - \\
      $ m^* $					    	 & 0.067  & 0.198 & $m_e$ \\
   $\rho $	    					 & 5300   & 2.3$\times 10^3$ & kg/m$^3$ \\
   $ c_t $					    	 & 2480   & 5$\times 10^3$ & m/s \\
   $ c_l $					   		 & 5290   & 9.15$\times 10^3$ & m/s \\
   $ \sigma_e $					   	 & 7   & - & eV \\
   $ h_{14} $					   	 & 1.4 $\times 10^{9}$   & 0 & eV/m \\
   $ l_0 $					    	 & $\sim$ 10   & $\sim$ 10   & nm \\
   $ l_d $					    	 & 0.63   & 0.25  & $\mu$m \\
   $ l_r $					    	 & 2.42   & 0.76 & $\mu$m \\
   $ \Xi_{d} $					   	 & -   & 5 & eV \\
   $ \Xi_{u} $					   	 & -   & 9 & eV \\
\hline
\end{tabular}
\caption
{ Typical bulk and quantum dot parameters for GaAs and Si/SiGe considered in this work. Values are taken from Refs.~\citenum{raith2011:PRB}, \citenum{raith2014:PSSB}.
}
\label{tab:parameters}
\end{table}

\bibliographystyle{unsrt}
\bibliography{References,quantum_dot}

\end{document}